# Access and *in situ* Growth of Phosphorene-Precursor Black Phosphorus

Marianne Köpf[1], Nadine Eckstein[1], Daniela Pfister[1], Carolin Grotz[1], Ilona Krüger[1], Magnus Greiwe[1], Thomas Hansen[2], Holger Kohlmann[3], Tom Nilges[1]*

1) Technische Universität München, Department of Chemistry, Lichtenbergstrasse 4
85748 Garching b. München, Germany

2) Institut Laue-Langevin, 71 avenue des Martyrs, CS 20156, 38042 Grenoble Cedex 9, France

3) Universität Leipzig, Fakultät für Chemie und Mineralogie, Institut für Anorganische Chemie, Johannisallee 29, 04103 Leipzig

Contact      Phone (TN): +49 89 289 13110
                E-mail: tom.nilges@lrz.tum.de

**Abstract**

Single crystals of orthorhombic black phosphorus can be grown by a short way transport reaction from red phosphorus and Sn/$SnI_4$ as mineralization additive. Sizes of several millimeters can be realized with high crystal quality and purity, making a large area preparation of single or multilayer phosphorene possible. An *in situ* neutron diffraction study has been performed addressing the formation of black phosphorus. Black phosphorus is formed directly via gas phase without the occurrence of any other intermediate phase. Crystal growth was initiated after cooling the starting materials down from elevated temperatures at 500 °C.

**Highlights:**

- Large black phosphorus crystals were grown by a short way transport reaction
- *In situ* neutron diffraction affirm the formation of black P directly from gas phase
- Large crystals can be used as starting material for phosphorene synthesis

## 1. Introduction

Phosphorene attracts great interest due to its superior physical properties and possible applications [1]. It has lately been synthesized from black phosphorus by exfoliation [2-5] or plasma assisted techniques [5; 6].

Recently, a plethora of exciting phosphorene properties and applications have been announced and reported [7]. Quantum-chemical calculations predicted enormous thermoelectric figures of merit [8; 9], superconductivity [10], optoelectronic features [11], Peierls-distortion [12], mechanical stability [13], and anisotropic features like electrical conductance [14] or layer-dependent band gap variation [15-17]. The calculated band gaps of phosphorene to bulk black phosphorus vary drastically from 1.54 eV for single layer phosphorene to 0.86 eV for a four layer arrangement [18]. Strain applied to phosphorene sheets results in anisotropic properties like a favored transport direction for charge carriers, the change from a direct to an indirect band transition or the modulation of the effective masses of electrons and holes [19].

Very promising reports are dealing with thin film solar cells [20], anisotropic optical and electrical properties [21] and the usage of phosphorene sheets in applications like field effect transistors [22; 23]. In all cases, the quality of the starting material black phosphorus and the possibility to tune, dope or modify the precursor plays a crucial role for the progress in phosphorene science.

Black phosphorus can also be prepared by various other methods like high-pressure synthesis [6; 24], recrystallization from Bi [24; 25] or Hg [25; 26] and by a complex melting and annealing sequence at elevated temperatures [26; 27]. All mentioned methods are often leading to small crystals with limited crystallinity, are time consuming, and – in some cases – they even involve toxic chemicals.

Orthorhombic black phosphorus is an allotrope, featuring layers of corrugated six-membered rings that are stacked along a crystallographic axis. Along the stacking direction van der Waals interactions are stabilizing the compound. Recently, van der Waals interactions have been identified as the most important attractive force determining the order of stability of all known phosphorus allotropes [28] substantiating black phosphorus to be the most stable allotrope at 0 K. At elevated temperature, fibrous phosphorus [29] is more stable than the black allotrope [30] but even here, van der Waals interactions between the layers in black phosphorus and between the tubes in fibrous phosphorus are the structure-determining forces. A structural section of black phosphorus and phosphorene is given in Figure 1.

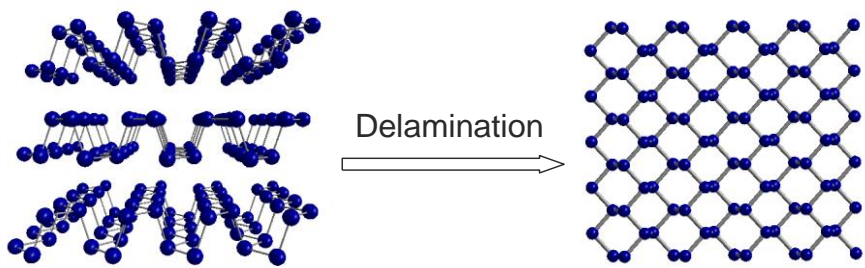

**Figure 1.** Structure sections of black phosphorus (left, side view onto sheets) and phosphorene (right, top view on single sheet).

The structural features of the black phosphorus/phosphorene couple and the phosphorene preparation are reminiscent to the only other stable 2D layered material, which also consists of a single element alone: graphene. Therefore, material scientists and physicists dealing with 2D materials are interested in the preparation of phosphorene from large black phosphorus crystals. Black phosphorus is stable on air for weeks [31] without any obvious surface reactions and even phosphorene can be handled on air within several hours to days [4]. Another field of interest is the chemical modification of black phosphorus or phosphorene by doping or chemical fragmentation [32-36]. Short way transport reactions are powerful tools to grow large single crystals with high crystal quality [37]. They also offer an access to perform doping via the gas phase during the crystal growth.

Herein we report on a modified black phosphorus synthesis based on a previously reported mineralizer-assisted short-way transport reaction [31; 38], resulting in short reaction times and large, high-quality crystals. Since *in situ* methods are a powerful tool to investigate phase formation and crystal growth [39], neutron diffraction *in situ* experiments have been performed at the ILL, Grenoble, to examine the phase formation and growing principle of black phosphorus.

## 2. Experimental Details

### 2.1 Synthesis of black phosphorus

Black phosphorus can be prepared from amorphous red phosphorus by a short way transport reaction using Sn and $SnI_4$ as reaction promoters or mineralization additives, $SnI_4$ was synthesized prior to usage as following [40]. Glacial acetic acid (25 mL, Merck, 99%) and acetic anhydride (25 mL, Acros Organics, 99+%) were placed in a round-bottomed flask. After Sn (0.5 g, Chempur, 99.999%) and $I_2$ (2 g, Chempur, 99.8%) were added, the mixture was heated to reflux until dissipation of Sn and an orange color of the solution indicated the reactions completion. Upon cooling the crude product crystallized as an orange solid, which was further purified by recrystallization from hot chloroform.

To prepare black phosphorus, Sn (20 mg, Chempur, 99.999%), $SnI_4$ (10 mg, synthesis procedure stated above) and red phosphorus (500 mg, Chempur, ultrahigh grade, 99.999+%,) were weighed in a silica glass ampoule of 10 cm length, an inner diameter of 1.0 cm and a wall thickness of 0.25 cm. The ampoule was evacuated and placed horizontally, with the starting materials mixture located at the hot end and the empty ampoule side towards the colder middle section of a Nabertherm L3/11/P320 muffle furnace, set to a temperature of 650 °C and then cooled during 7.5 h to 550 °C. Heating elements are located within the walls of the furnace.

In contrast to our previously reported preparation method [31] we were able to reduce the amount of side phases drastically. So far, Sn, $SnI_4$, and Au were needed to be mixed with red phosphorus with the consequence that side phases like $Au_3SnP_7$, AuSn, or $Au_2P_3$ were formed, in addition to the main product. Only tin and iodine from $SnI_4$ are still potential candidates to form side products after the removal of Au from the synthesis protocol.

*2.2 Characterization*

For determination of the sample composition, a fraction of as prepared crystals was dissolved in $HNO_3$ at 120 °C and an ICP-OES measurement was performed using an Agilent Technologies ICP-OES 725 device.

To gain crystallographic data powder X-ray diffraction was done using a Stoe STADI P, Cu source, $CuK\alpha_1$ radiation and a Dectris Mythen 1K detector.

Raman spectroscopy was carried out using a Senterra Spectrometer of Bruker Optics GmbH equipped with a 785 nm laser and a power of 10 mW. The average data was gained of 5 separate measurements with 5 s of integration time each and a magnification of 50x.

*In situ* neutron powder diffraction data were collected at the high intensity 2-axis diffractometer D20 [41] at the Institut Laue-Langevin (ILL) in Grenoble. The instrument was set at a take-off angle of 118° $2\theta$ ("high resolution"), which gives a wavelength of $\lambda \approx 1.87$ Å when using the (115) reflection of its vertically focusing germanium monochromator. No further Soller collimator has been used to reduce the natural primary divergence of $\alpha_1 \approx 27'$. The monochromatic neutron beam at the sample has been set by $B_4C$ absorbing slits inside the furnace (see below) to be 4 cm high (plus 2 × 5 mm penumbra). Absorbing slits, at about 50 cm upstream the sample position, define the horizontal beam width not to exceed the sample diameter by more than a penumbra of 2 × 5 mm. The sample inside a sealed quartz ampoule of 8 mm inner diameter has been placed in a cylindrical sample holder of a slightly bigger diameter and 6 cm height, made of thin (40 µm) vanadium foil, which has been centered in a 30 mm diameter resistive heating element of 100 µm vanadium foil of 20 cm height inside a vacuum vessel of 60 cm diameter. The large diameter allows for in-device absorbing $B_4C$ sheets, in particular a direct beam stop, in order not to collect parasitic neutron diffraction peaks from the vessel material in the open position sensitive detector, as no radial oscillating collimator has been used, which could suppress parasitic scattering contributions from material far enough away from the centered sample position. The furnace allowed for controlled heating on a linear temperature ramp of 200 K/h from room temperature up to 650°C, holding temperature for 30 min, cooling on a linear ramp to 500°C, holding temperature for 1 h and cooling on a linear ramp down to 156°C, where the cooling rate eventually becomes slower than the cooling rate of 100 K/h. The diffracted neutrons in the diffraction plane are counted by a large, one-dimensional curved position sensitive detector, which covers 153.6° in $2\theta$ in steps of 0.1°. Diffraction patterns have been collected during the heating and cooling cycle for 5 min every 5 min (dead-time between two acquisitions negligible, about 100 ms). The raw data has been corrected for relative detector efficiency (as each of the 1536 detector cells may have a slightly – by about 1% – different counting efficiency) and exported in the adequate data format for further analysis.

### 3. Results and Discussion

*3.1 Short-way transport growth of black phosphorus*

In order to allow an effective and fast synthesis of black phosphorus we have tried to optimize the reaction conditions in such a way that main product and side phases are well separated from each other.

Sn and SnI$_4$ will comproportionate lower than the reaction temperature. We use an excess of Sn to allow a quantitative transformation of SnI$_4$ to SnI$_2$ and to avoid SnI$_4$ decomposition and thus formation of I$_2$. The remaining Sn reacts with phosphorus from the gas phase to Sn-phosphides. An intrinsically present temperature gradient of 45 to 50 °C is applied to the whole ampoule and black phosphorus is grown within this temperature gradient. Large bunches of black phosphorus are formed as parallel arranged black phosphorus crystals (see Figure 2).

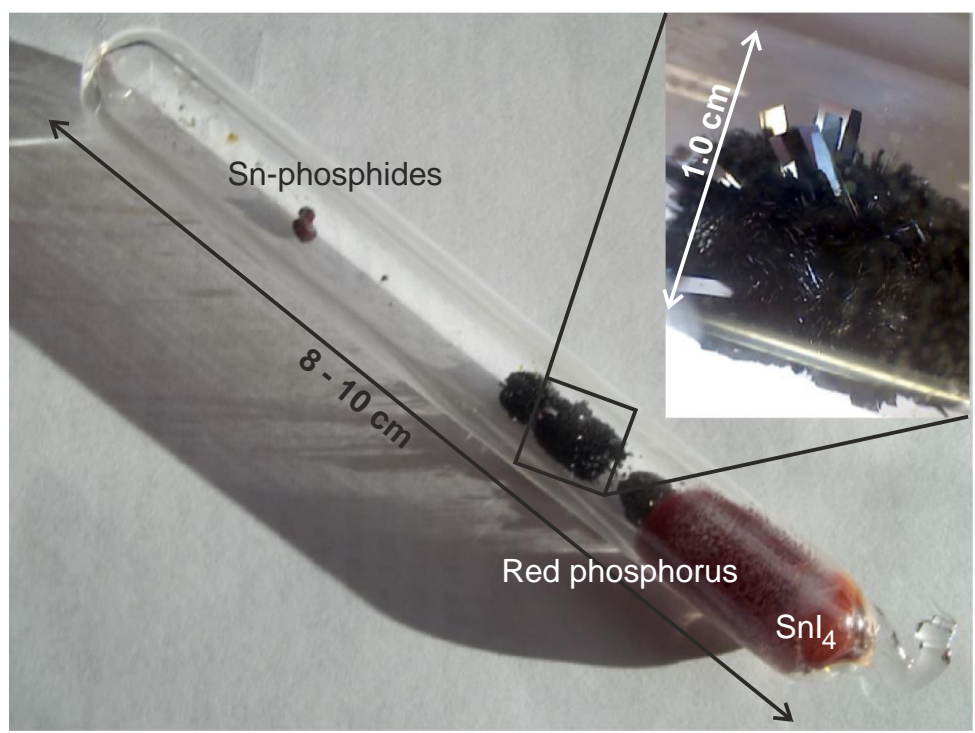

**Figure 2.** A representative silica glass ampoule after the synthesis of black phosphorus. SnI$_4$ (orange) and red phosphorus (red) from the gas phase are condensed at the right hand side of the ampoule. Black phosphorus has been formed in large bunches. Excessive Sn has reacted to Sn-Phosphides, which are present in small round spheres.

Upon cooling, SnI$_2$ will disproportionate to SnI$_4$ and Sn within the ampoule again. Sn seemed to have reacted almost completely to Sn-phosphides and molecular SnI$_4$ which condensates mostly in the cold section of the ampoule (see Figure 2). Small portions of SnI$_4$ which might be condensed onto the black phosphorus crystals can be removed by refluxing the product in boiling toluene until the toluene remains clear (ultrasonic bath 20-45 min). An ICP-OES experiment on as-prepared crystals has been performed leading to a purity of > 99.9 at-%. We found no significant amounts of any other element than phosphorus (e.g. tin, iodine, or silicon from the ampoule material) in the final product.

A fraction of the final product was ground into small crystals and measured by powder X-ray diffraction. No additional reflections to those expected for black phosphorus were detected. Texturing occurred due to the extreme layered morphology of black phosphorus and the pre-orientation of crystals during the experiment. For Rietveld refinement (see Figure 3), a preferred orientation in (010) was applied according to the March-Dollase-Model implemented in the Jana2006 program [42].

The identity of the synthesized black phosphorus was also proven by Raman spectroscopy. The observed modes $A_{1g}$, $B_{2g}$ and $A_{2g}$ (see Figure 4) are fully consistent with literature [43; 44].

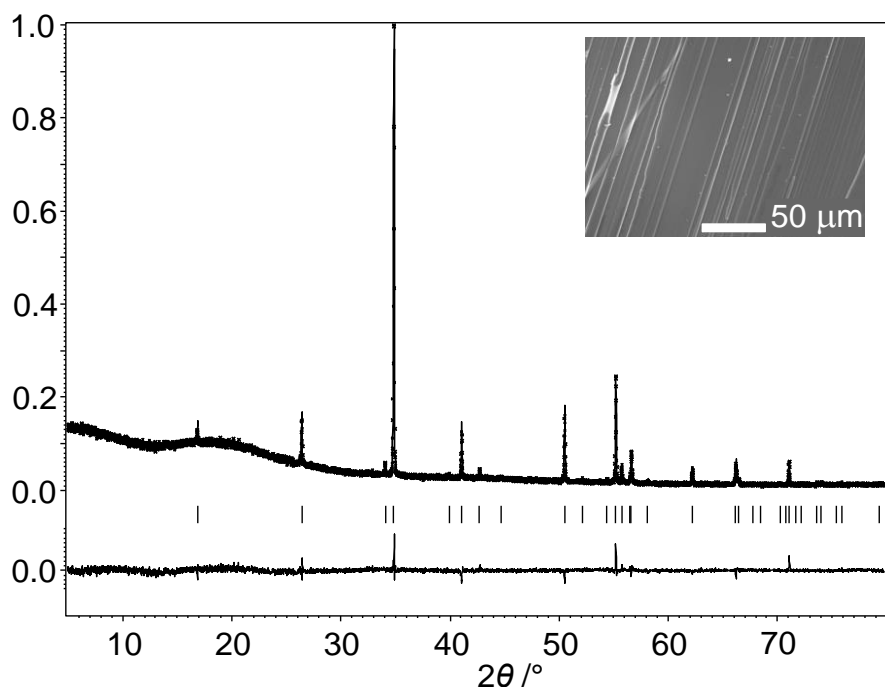

**Figure 3.** Powder X-ray diffractogram and Rietveld plot of black phosphorus with a starting structure model in space group *Cmce* taken from [31]. No crystalline impurity was observed. Lattice parameters are $a$ = 3.3264(2); $b$ = 10.5194(8), and $c$ = 4.3937(2) Å. Hence, they are slightly higher than the ones reported ($a$ = 3.3164(1); $b$ = 10.484(2), and $c$ = 4.379(1) Å) [31]. A representative SEM picture of black phosphorus and a resulting Rietveld refinement including a difference pattern are given. Phosphorus on 8$f$ (0 0.0970(4) 0.0871(4)), $U_{iso}$ = 0.016(2) Å$^2$, w$R2$ = 0.0798, $R_p$ = 0.0542.

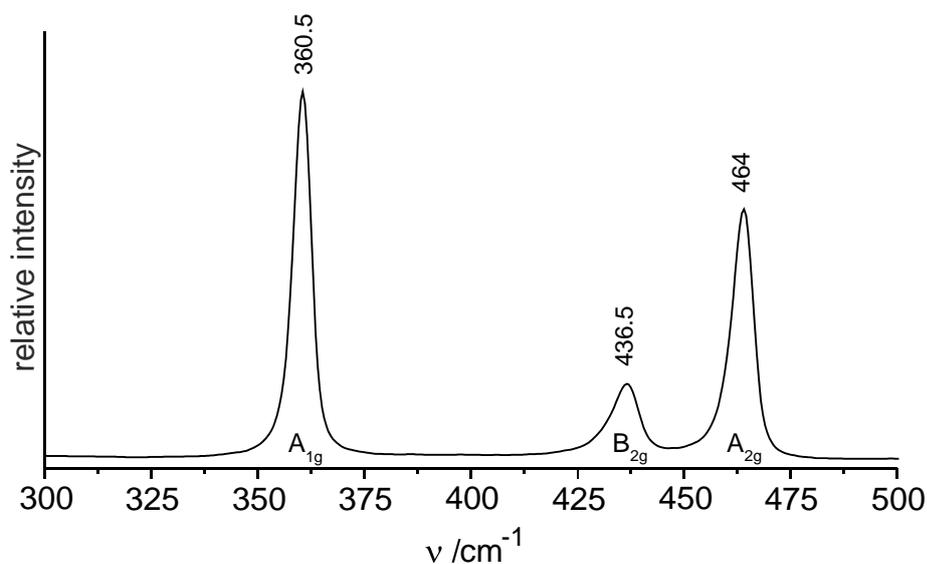

**Figure 4.** Raman spectrum of black phosphorus. The Measurement was taken at ambient conditions with a 785 nm laser. Peak assignment was done according to literature [43; 44].

*3.2 In situ growth of black phosphorus*

By *in situ* neutron diffraction experiments we evaluated the formation and growth of black phosphorus by the mineralization procedure stated herein. Due to safety reasons the overall phosphorus loading was reduced to 45 mg and the ampoule shortened to 25 mm with an inner diameter of 8 mm. Fortunately, black phosphorus shows such a reasonable crystallinity that diffraction patterns taken in 5 minutes showed a sufficient counting statistics despite the small sample size as compared to the usual scale of neutron diffraction experiments. After heating up to 650°C and subsequent cooling down to 500°C and holding temperature there (for one hour), the black phosphorus started to crystallize with an induction period of about 50 min, just 10 min before the temperature was lowered again. (Details are shown in Figure 5). Upon cooling at a rate of 100 K/h, black phosphorus crystallized continuously during this process. Only reflections of black phosphorus are visible in the diffraction pattern. Due to the growth of black phosphorus onto the ampoule walls texture effects occur, resulting in weak or non-observable intensities for some reflections. The final neutron pattern at 156°C is also shown in Figure 5.

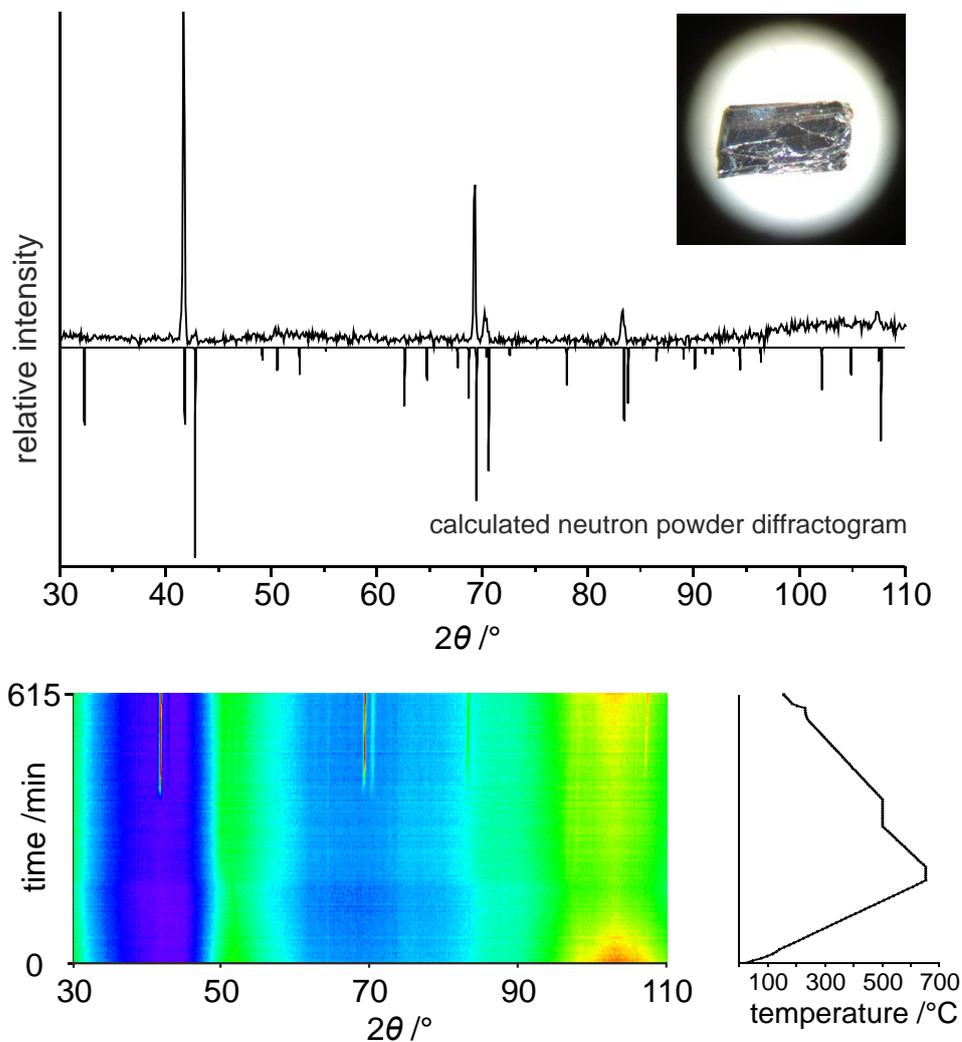

**Figure 5.** Top part: Neutron diffraction pattern of black phosphorus, taken at 156°C, showing the main reflections of black phosphorus. A picture of a separated such grown single crystal is given as an inset. Bottom part: 2D-representation of the temperature-dependent *in situ* neutron diffraction experiment together with the applied temperature program, both correlated to the experiments time (total 615 min). Black phosphorus is the only crystalline species formed during the experiment. Scattering intensities in the false-color picture last from low intensities in blue via green to high intensities in orange. A strong background was present due to the usage of silica glass ampoules during the experiment.

Crystal growth of black phosphorus can be followed by the evolution of the first strong reflection (040) at 41.7° $2\theta$ occurred in the Neutron diffraction pattern (see Figure 6). After the observation of the first Bragg intensity at about 365 min – time measurement started at the point where heating was applied – half of the integrated intensity of this reflection was reached by additional 40 min, or 405 min total reaction time. This finding directly leads to two main conclusions. The first conclusion is that black phosphorus is the only crystalline phase observed, and no crystalline side phase was detected. Secondly, the formation and growth of black phosphorus is fast (within minutes) after initial nuclei have been formed. Therefore, a certain induction period is needed where no information can be derived from this experiment. In order to evaluate the kinetics of this crystal growth, we plan to perform isothermal *in situ* experiments in the near future.

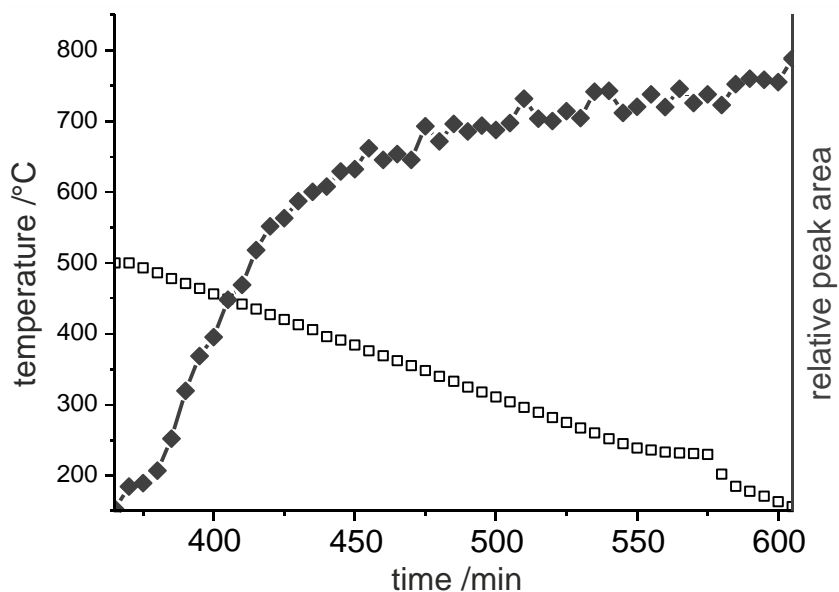

**Figure 6.** Temperature (□) and integrated intensity (◆) of the (040) reflection at 41.7° ($2\theta$) of black phosphorus in correlation with experimental time.

### 3. Conclusion

We were able to significantly improve the synthesis procedure for black phosphorus and the crystal quality in comparison with our earlier reports. Expensive Au was rejected from the synthesis procedure completely, thus reducing costs and the amount of side phases drastically. After optimizing the temperature program, no significant condensation of red phosphorus and $SnI_4$ has been observed on the black phosphorus crystals. During the former procedure of synthesis, spots of $SnI_4$ were present in the black phosphorus bunches. Furthermore, the formation area of black phosphorus and the condensation of side phases were overlapping. Dependent on the demanded grade of purity (amounts smaller than 0.01 at-% of side products can still be present on the black phosphorus crystals) the final product can be used without any further purification steps.

Using this improved method orthorhombic black phosphorus can now be prepared by short-way transport reaction form red phosphorus applying a mineralization approach. Sn and $SnI_4$ are added in small amounts to assist the formation and crystal growth. Purity and phase analysis have been checked by ICP-OES and X-ray powder diffraction, respectively. This procedure allows a cheap, fast, and effective growth of black phosphorus single crystals in high quality, large amounts, and sizes up to several millimeters. Single crystals of several millimeters lengths might be useful for the effective preparation of single or multi-layer, large area phosphorene.

## 4. Acknowledgement

This work is financed by the DFG via grant no. Ni1095/2-2, within the Priority research program 1415. We thank the ILL Grenoble and D20 team for their support as well as Prof. Dr. Sonja Berensmeier and M. Sc. Sebastian Schwamminger, TUM, Garching, Germany for Raman measurement. We gratefully thank M. Sc. Jennifer Ludwig for proof-reading of this manuscript.